\documentclass{pasj01}
\Received{2019 August 13}
\Accepted{2019 October 17}
\Published{$\langle$publication date$\rangle$}
\usepackage{scalefnt}
\usepackage{url}
\usepackage{natbib}

\usepackage{ulem}

\begin{document}

\title{Vertical Structure and Kinematics of the Galactic Outer Disk}
\author{
Nobuyuki \textsc{Sakai}\altaffilmark{1},
Takumi \textsc{Nagayama}\altaffilmark{2},
Hiroyuki \textsc{Nakanishi}\altaffilmark{3},
Nagito \textsc{Koide}\altaffilmark{3},
Tomoharu \textsc{Kurayama}\altaffilmark{4},
Natsuko \textsc{Izumi}\altaffilmark{5},
Tomoya \textsc{Hirota}\altaffilmark{6, 7},
Toshihiro \textsc{Yoshida}\altaffilmark{2},
Katsunori M. \textsc{Shibata}\altaffilmark{6, 7}, and
Mareki \textsc{Honma}\altaffilmark{2, 7}.
}%
\altaffiltext{1}{Korea Astronomy $\&$ Space Science Institute, 776, Daedeokdae-ro, Yuseong-gu, Daejeon 34055, Republic of Korea}
 \altaffiltext{2}{Mizusawa VLBI Observatory, National Astronomical
   Observatory of Japan, Hoshigaoka 2-12, Mizusawa, Oshu, Iwate 023-0861, Japan}
 \altaffiltext{3}{Graduate School of Science and Engineering, Kagoshima
   University, 1-21-35 Korimoto, Kagoshima 890-8580, Japan}
 \altaffiltext{4}{Teikyo University of Science, 2-2-1 Sakuragi, Senjyu, Adachi-ku, Tokyo, Japan}
\altaffiltext{5}{College of Science, Ibaraki University, 2-1-1 Bunkyo, Mito, Ibaraki 310-8512, Japan}
\altaffiltext{6}{Mizusawa VLBI Observatory, National Astronomical
   Observatory of Japan, Osawa 2-21-1, Mitaka, Tokyo 181-8588, Japan}
 \altaffiltext{7}{Department of Astronomical Science, The Graduate University for Advanced Studies (SOKENDAI), Osawa 2-21-1, Mitaka, Tokyo 181-8588, Japan}

 \email{nsakai@kasi.re.kr}


\maketitle

\begin{abstract}
We report measurements of parallax and proper motion for four 22 GHz water maser sources as part of VERA Outer Rotation Curve project. All sources show Galactic latitudes of $>$ 2$^{\circ}$ and Galactocentric distances of $>$ 11 kpc at the Galactic longitude range of 95$^{\circ}$ $< l <$ 126$^{\circ}$. The sources trace the Galactic warp reaching to 200$\sim$400 pc, and indicate the signature of the warp to 600 pc toward the north Galactic pole. The new results along with previous results in the literature show the maximum height of the Galactic warp is increased with Galactocentric distance. Also, we examined velocities perpendicular to the disk for the sample, and found an oscillatory behavior between the vertical velocities and Galactic heights. This behavior suggests the existence of the bending (vertical density) waves, possibly induced by a perturbing satellite (e.g. passage of the Sagittarius dwarf galaxy). 
\end{abstract}

\section{Introduction}\label{Intro}
The structure of the Galactic disk has been observationally examined since 1950s at multi-wavelengths (e.g., \citet{1957BAN....13..201W} at radio; \citet{1976A&A....49...57G} at radio and optical; \citet{2001ApJ...556..181D} at infrared; \citet{2003A&A...397..133R} at radio and optical; \citet{2014A&A...569A.125H} at radio and optical), and owing to difficulty of accurate distance measurement this type of study is still an important and challenging topic in the Galactic astronomy. Since the 2000s, trigonometric parallax measurements using the Very Long Baseline Interferometry (VLBI) technique have been intensively conducted in the Galaxy for better understanding the structure and kinematics of the Milky Way by VERA (\citealt{2012PASJ...64..136H}), VLBA BeSSeL Survey (\citealp{2009ApJ...700..137R, 2014ApJ...783..130R, 2019arXiv191003357R}), EVN (e.g., \citealp{2010A&A...511A...2R, 2012A&A...539A..79R}), and LBA (e.g., \citealp{2015ApJ...805..129K, 2017MNRAS.465.1095K}). \citet{2019arXiv191003357R} compiled $\sim$200 parallaxes for Galactic star-forming regions, and derived number, widths, pitch angles of the Galactic spiral arms as well  as the Galactic constants, the solar location perpendicular to the disk (i.e., $z_{\odot}$), and a model of Galactic rotation curve. However, there has been a lack of observational study in a part of the Milky Way. For instance, \citet{2019arXiv191003357R} listed $\sim$200 parallax results of which only 11 sources were available for the Outer arm (see Table 1 of \citealp{2019arXiv191003357R}). This is due to low star formation rate and general weakness of masers at the outer disk (e.g., \citealp{1993A&AS...98..589W}; \citealp{2000A&A...360..311S}). 

H {\scriptsize I} (21 cm) line data has been widely used for studying the structure of the Galactic disk. The Galactic H {\scriptsize I} gas layer is known to be exceedingly flat within a Galactocentric distance ($R$) of 7 kpc (\citealp{1960MNRAS.121..132G}; \citealp{1960MNRAS.121..123B}). Beyond the Galactocentric distance of  7 kpc, prominent structure of the H {\scriptsize I} warp, regarded as the deviation of the center of the H {\scriptsize I} gas layer from the Galactic mid-plane, has been known and studied since 1957 (e.g., \citealp{{1957AJ.....62...90B}}; \citealp{{1957AJ.....62...93K}}; \citealp{{1957BAN....13..201W}}; \citealp{1982ApJ...259L..63K}; \citealp{2003PASJ...55..191N}; \citealp{2006ApJ...643..881L}; \citealp{2007A&A...469..511K}; \citealp{2016PASJ...68....5N}). The northern Galactic disk is warped toward the north Galactic pole, while the southern Galactic disk is warped toward the south Galactic pole. The warp's height perpendicular to the disk is monotonically increased with Galactocentric distance (e.g., see Fig. 1 of \citealp{1960MNRAS.121..132G}; Fig. 2 of \citealp{1986A&A...163...43S}; Fig. 12 of \citealp{2006PASJ...58..847N}; Fig. 16 of \citealp{2007A&A...469..511K}). Indeed, Figure 2 of \citet{2006ApJ...643..881L} clearly showed that the Galactic H {\scriptsize I} outer disk is warped toward the north Galactic pole at the Galactic longitude range of $\sim$30$^{\circ}$ $< l <$ $\sim$150$^{\circ}$ and $\sim$10 $< R <$ $\sim$30 kpc. 

Using five VLBI parallax results for the Outer arm, \citet{2015APJ...800....2H} showed clear evidence of warping reaching to $z$ = 200$\sim$400 pc at 75$^{\circ}$ $< l <$ 135$^{\circ}$ and 10.8 $< R <$ 13.2 kpc, although the number of the observational results was limited. Using $Gaia$ DR2 results for upper main sequence (i.e. young) stars and giants, \citet{2018MNRAS.481L..21P} detected kinematic signature of the Galactic warp, for which large-scale systematic motion perpendicular to the disk is expected at the Galactic anticenter (see Fig. 3 of \citealp{2018MNRAS.481L..21P}). However, over densities of the young stars were only detected in the nearby spiral arms (i.e. Perseus, Local and Sagittarius arms). Thus, the structure of the Outer arm is poorly understood at this moment.

While astrometric observations for the Galactic warp are starting to become significant as mentioned above, one can refer to simulation and theoretical studies for understanding the origin of the Galactic warp (e.g., \citealp{1969ApJ...155..747H}; \citealp{1992ARA&A..30...51B}; \citealp{2006MNRAS.370....2S}; \citealp{2008gady.book.....B}; \citealp{2011Natur.477..301P}; \citealp{2013MNRAS.429..159G}). For instance, \citet{2013MNRAS.429..159G} conducted simulation study to examine the effect of the passage of the Sagittarius dwarf galaxy into the Galactic disk. It was revealed that both the height of the Galactic warp and velocity perpendicular to the disk vary sinusoidally in radial direction. Also, an oscillatory behavior between the both sinusoids (i.e. a phase offset of $\pi$/2 (rad) between the both sinusoids) was discovered. Validation of the simulations results with astrometric data is important for better understanding the Galactic warp.

Here, we report parallax and proper motion results for the Outer arm as part of VERA Outer Rotation Curve (ORC) project (see the detail of the project in \citealp{2012PASJ...64..108S}), which more clearly reveal the structure and kinematics of the Outer arm as well as the Galactic warp. In section 2, we describe the observational setup. In section 3, we outline the data reduction. In section 4, we show the observational results. Using our results along with previous results in the literature, we discuss three-dimensional (3D) structure and vertical velocities of the Galactic warp at $R >$ 7 kpc, and compare those with a Galactic warp model in section 5. In section 6, we summarize the paper.

\begin{table*}[htbp] 
\caption{Source and Observation Information.} 
\label{table:4} 
\scalebox{0.92}[0.92]{
\begin{tabular}{lllrclllr}
\hline 
\hline 
Source&R.A. &Decl. 	  &Obs. Date	&&Source &R.A. &Decl.&Obs. Date \\
	   &hh:mm:ss	     &$^{\circ}$: ': "		  & in UT&&&hh:mm:ss&$^{\circ}$: ': "& in UT\\
	       &(J2000) &(J2000) &20yy/mm/dd&&&(J2000)&(J2000) &20yy/mm/dd\\ 
\hline
G095.05+3.97	&21:15:55.6798    	&+54:43:31.328&&&G097.53+3.18&21:32:12.4400  	&+55:53:49.600&A. 12/01/18	 \\            
 J2123+5500	&21:23:05.3135    	&+55:00:27.325&&&J2123+5500	&21:23:05.3135    	&+55:00:27.325&B. 12/03/20	\\       
 			\multicolumn{3}{l}{(SA = 1.1$^{\circ}$; PA = 73.9$^{\circ}$)\footnotemark[$*$]}&C. 12/05/21&&\multicolumn{3}{l}{(SA = 1.6$^{\circ}$; PA = -123.6$^{\circ}$)\footnotemark[$*$]}&C. 12/05/21	\\
 	 &	&&D. 12/08/20&&		&&&D. 12/08/20\\       
 	&  	&&E. 13/02/02&&  	&&&E. 13/02/02\\  
 	&  	&&F. 13/04/05&&		&&&F. 13/04/05\\ 
 	&  	&&G. 13/11/01&&  	&&&G. 13/11/01\\ 
 	&  	&&H. 14/01/29&&		&&&H. 14/01/29\\
 	&  	&&I. 14/04/21&&		&&&I. 14/04/21\\ 	%
	&  	&&J. 14/09/04&&		&&&J. 14/09/04\\       
 	&  	&&K. 14/11/07&&		&&&K. 14/11/07	\\ 
&  	&&L. 14/12/22\\      
&  	&&M. 15/04/12\\
\\       

G102.35+3.64	&21:57:25.1841    	&+59:21:56.614&N. 12/08/19&&G125.52+2.03&01:15:40.8027  	&+64:46:40.766&a. 12/01/20	 \\            
 J2148+6107	&21:48:16.0422    	&+61:07:05.794&O. 13/01/26&&J0128+6306	&01:28:30.5650    	&+63:06:29.882&b. 12/03/27	\\       
 	\multicolumn{3}{l}{(SA = 2.1$^{\circ}$; PA = -129.5$^{\circ}$)\footnotemark[$*$]}&P. 13/04/02&&\multicolumn{3}{l}{(SA = 2.2$^{\circ}$; PA = 138.4$^{\circ}$)\footnotemark[$*$]}&c. 12/05/14	\\
 	&  	&&Q. 13/08/23&&		&&&d. 12/08/21\\       
 	&  	&&R. 13/11/30&&  	&&&e. 13/01/25\\  
 	&  	&&S. 14/03/21&&		&&&f. 13/03/10\\ 
 	&  	&&T. 14/05/23&&  	&&&g. 13/08/15\\ 
 	&  	&&U. 14/09/01&&		&&&h. 13/10/19\\
 	&  	&&V. 14/11/14&&		&&&i. 14/03/23\\ 	%
	&  	&&W. 15/02/09&&		&&&j. 14/05/24\\       
 	&  	&&X. 15/05/06&&		&&&k. 14/09/02	\\ 
	&  	&&Y. 15/12/18&&		&&&l. 14/11/10\\      
\hline 

\multicolumn{4}{@{}l@{}}{\hbox to 0pt{\parbox{185mm}{
\par\noindent
\\
\small
Column 1 lists an observed
22 GHz H$_{2}$O maser source (as denoted by ``G") and a background
QSO (as denoted by ``J"). Columns 2-3 represent equatorial coordinates for the source in (J2000). Column 4 displays date of the observation.
Columns 5-8 are the same as the columns 1-4. \\
\footnotemark[$*$]Parenthesis represents Separation Angle (SA) and Position Angle (PA) East of North of the background QSO with respect to the maser source.
}\hss}}
\end{tabular} 
}                                        
\end{table*}

\section{Observation}

As part of VERA ORC project, we carried out VERA astrometric observations between January 18th, 2012 and December 18th, 2015 for four H$_{2}$O maser sources at a rest frequency of 22.235080 GHz (see Table \ref{table:4}). The procedure of the observations is the same as our previous observations (\citealp{2012PASJ...64..108S, 2015PASJ...67...69S}; \citealp{2015PASJ...67...68N}; \citealp{2019arXiv190909930K}). Each source and an adjacent background QSO were observed simultaneously with the VERA dual-beam system \citep{2000SPIE.4015..544K} for compensating atmospheric phase fluctuation. Also, a bright continuum source was observed $\sim$5 (min) every 80 (min) for clock offsets calibration of VERA antennas in each observation.

During the observations, left-handed circular polarization data were recorded at 1,024 Mbps with 2-bit quantization after filtering performed with the VERA digital filter \citep{2005PASJ...57..259I}. The data sets were correlated with Mitaka FX correlator \citep{1998ASPC..144..413S}\footnote{Please
also see the VERA HP: \\ \url{http://www.miz.nao.ac.jp/en/content/facility/vera-correlation-center}}, except for the last observational data (as denoted by ``Y'' in Table \ref{table:4}). The last data was correlated with Mizusawa software correlator because Mitaka FX was replaced with the software correlator in 2015 summer\footnote{\url{https://www.miz.nao.ac.jp/veraserver/system/fxcorr-e.html}}.

For the former data sets, one IF of 16 MHz band was assigned for the maser and correlated with 512 channels, giving a frequency (velocity) spacing of 31.25 kHz (0.42 km s$^{-1}$) at the rest frequency. For continuum sources, 224 MHz band consisting of 14 IFs were assigned and correlated with 64 channels for each IF. Regarding the latter data, the same velocity spacing of the former data was applied for the maser source, while the continuum data consisted of 240 MHz/ 15 IFs and was correlated with 32 channels for each IF. 

\section{Data reduction}\label{sec:data}

Data reduction was conducted with the NRAO Astronomical Image Processing System ($AIPS$; \citealt{1996ASPC..101...37V}). We employed general procedure as described in \citet{2015PASJ...67...69S}. Basically, each extragalactic continuum source was used as a phase-reference source, and target maser was imaged by the phase referencing to measure (relative) positional offset as a function of time for 2$\sim$3 years. The position offsets of masers, relative to continuum sources, were used to measure trigonometric parallaxes and proper motion components for individual maser sources. Note that in case of G102.35+3.64 the continuum source J2148+6107 was slightly faint ($\sim$ 50 mJy) during the observations, and thus we used the maser as a phase reference and the continuum source was imaged by the phase referencing.

For the parallax measurements, we selected compact maser spots which showed image SNR (Signal to Noise Ratio) of $>$ 10 and survived more than 1 year. This is because parallax and proper motion components are highly correlated less than 1 year. Since each maser feature, consisting of maser spots with different $V_{{\rm LSR}}$, had a velocity width of few km s$^{-1}$ and showed shift of peak intensity in the velocity range during the observations, we regarded the brightest position of the maser as the centroid position and used it for the parallax determination. Experimentally, we have confirmed this method works well for the VERA data.  

For the proper motion measurements, we selected maser features which showed image SNR of $>$ 10 and were detected more than two epochs so as to avoid misconnection of different maser feature's positions.

\begin{table*}[tbp]
\caption{Observational results.}
\label{table:1}
\begin{center}
\begin{tabular}{llcccrcc}
\hline
\hline
Target	&IRAS	&Parallax ($\pi$)&Distance &$\mu_{\alpha} \rm{cos}\delta$	&$\mu_{\delta}$ \ \ \ \ \ \ &Noise&S/N	 \\

	&&(mas)&(kpc)&(mas yr$^{-1}$)	&(mas yr$^{-1}$)	&(mJy beam$^{-1}$)				\\
  \hline
G095.05+03.97		&21144+5430			&0.108$\pm$0.023	&9.26$^{+2.51}_{-1.63}$&$-$2.44$\pm$0.21	 &$-$2.63$\pm$0.17&355	&34\\
G097.53+03.18		&21306+5540			&0.177$\pm$0.028	&5.65$^{+1.06}_{-0.77}$&$-$2.64$\pm$0.20	 	&$-$2.38$\pm$0.22&108&32\\
G102.35+03.64\footnotemark[$*$]		&21558+5907		&0.154$\pm$0.021	&6.49$^{+1.03}_{-0.78}$&$-$2.53$\pm$0.33	 	&$-$2.14$\pm$0.33&\ \ \ \ \ \ 3\footnotemark[$*$]&23	\\
G125.52+02.03		&01123+6430		&0.145$\pm$0.023		&6.90$^{+1.30}_{-0.94}$&$-$1.20$\pm$0.21	&$-$0.33$\pm$0.27	&\ \ 43&44	\\

\hline

\multicolumn{5}{@{}l@{}}{\hbox to 0pt{\parbox{155mm}{\footnotesize
\par\noindent \\
Columns 1-2 show target and IRAS names, respectively. Columns 3 and 4 represent parallax and corresponding distance (i.e. 1/$\pi$) results. Columns 5-6 display proper motion components in east ($\alpha$cos$\sigma$) and north ($\sigma$) directions, respectively. Columns 7-8 show typical values of rms noise and signal to noise ratio on the phase-referenced image. \\
\footnotemark[$*$] The maser source was used as a phase reference and the back ground QSO was imaged by the phase referencing (see the text).

}\hss}}

\end{tabular}
\end{center}
\end{table*}

\begin{figure*}[tbhp] 
 \begin{center} 
     \includegraphics[scale=1.3]{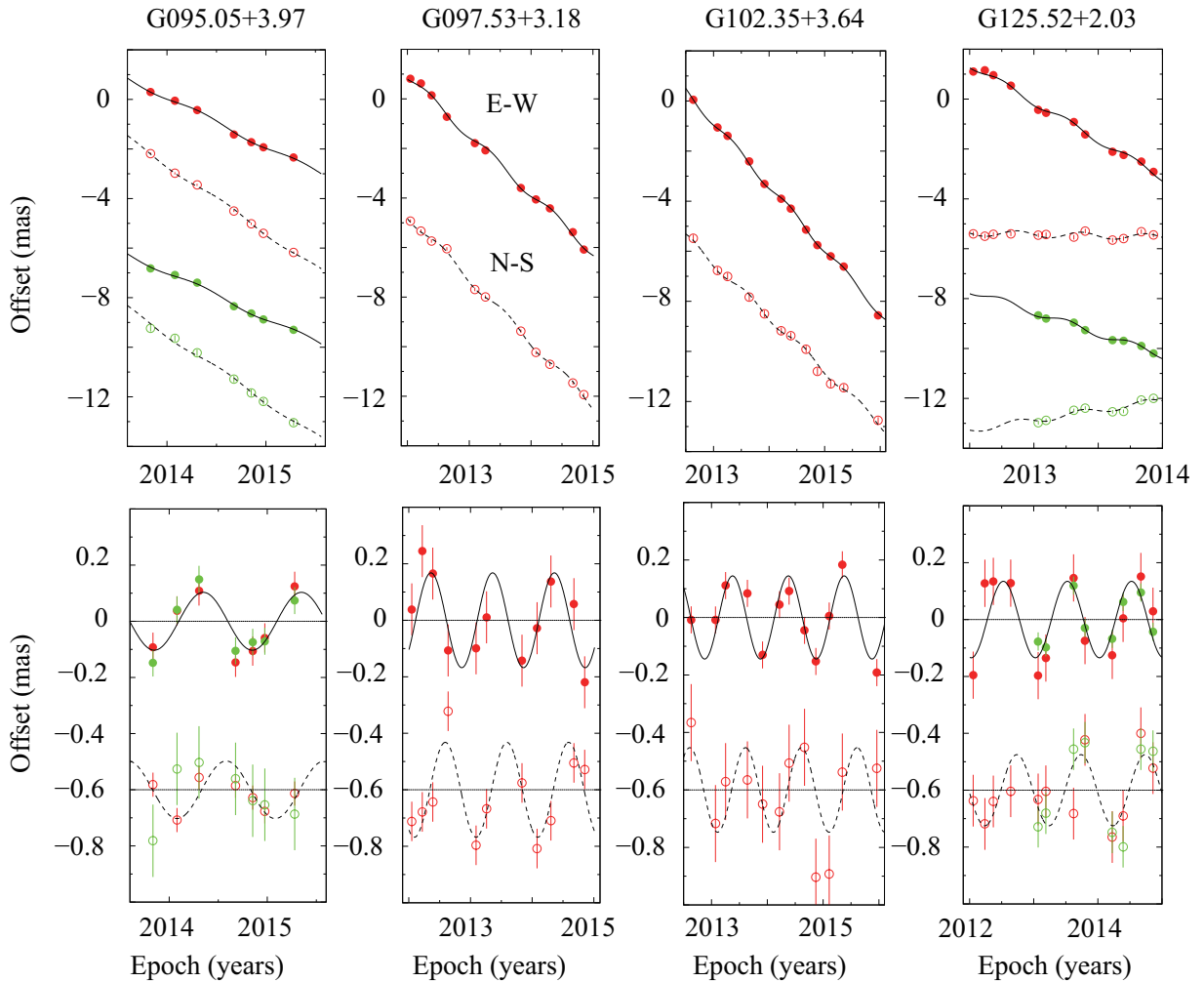} 
\end{center} 
\caption{ Results of parallax and proper-motion 
fitting. Plotted are position offsets of maser spots with respect to background
QSOs toward the east and north as a function of time. For clarity, the data toward the north direction is plotted offset from the
 data toward the east direction. \textit{(Top row)} The best-fit models in the east and north directions are shown
as continuous and dashed curves, respectively. Red and green circles indicate 1st and 2nd maser features, respectively (see Table \ref{table2} of Appendix \ref{appendix:1}). 
\textit{(Bottom row)} Same as top row, but with proper motions removed.}
\label{fig:1} 
\end{figure*}  

\section{Results}

We provide 22 GHz parallax and proper motion results for four maser sources (see Table \ref{table:1} and Fig. \ref{fig:1}). Supplemental information of the results are summarized in Appendix \ref{appendix:1}. To estimate proper motion of the central star that excites the maser, we averaged proper motion values for all maser features. For the uncertainty of the estimation, we added $\pm$5 km s$^{-1}$ in quadrature to the statistical error for each motion component in Table \ref{table:1}. The uncertainty of 5 km s$^{-1}$ is consistent with a velocity dispersion of typical molecular cloud (e.g. \citealp{1985ApJ...295..422C}). For G102.35+03.64, only single proper motion value was measured (see Table \ref{table3} of Appendix \ref{appendix:1}) and thus to be conservative we added $\pm$10 km s$^{-1}$ in quadrature to the statistical error in Table \ref{table:1}. We emphasize that averaged $V_{{\rm LSR}}$ values of the maser features for individual sources are consistent with $V_{{\rm LSR}}$ values, estimated from observations of thermal line emissions (e.g., CO) from the parent clouds, within errors. It indicates that averaged proper motion values display the proper motions of the central stars reasonably for individual sources.

Since parallax and proper motion values were previously reported for some sources, we compare our results with the previous ones in the following subsections.

\subsection{G097.53+03.18}
Using VLBA, \citet{2014ApJ...783..130R} reported a 22 GHz parallax of 0.133$\pm$0.017 mas for the source. The parallax difference between \citet{2014ApJ...783..130R} and this paper is 1.3$\sigma$ and statistically insignificant. The independent results could reasonably have come from random differences and we combine the both results by variance-weighting to give a best parallax for G097.53+03.18 of 0.145$\pm$0.015 mas.

The differences of proper motion components between the both results are less than 1$\sigma$ for each motion component. Thus, we assign the best proper motion for the source of ($\mu_\alpha$cos$\sigma$, $\mu_{\delta}$) = (-2.74$\pm$0.16, $-$2.41$\pm$0.16) mas yr$^{-1}$ by the variance weighting.

\subsection{G125.52+02.03}
Using VERA, \citet{2019arXiv190909930K} reported a 22 GHz parallax of 0.151$\pm$0.042 mas, where the same data sets were used, except for the first epoch (as denoted by ``a'' in Table \ref{table:4}). The result of \citet{2019arXiv190909930K} is consistent with our result, but this paper succeeded to reduce the parallax error from 42 $\mu$as to 23 $\mu$as. Since there are differences between the both papers on data reductions, we checked each difference if our parallax error is improved or degraded by applying each difference. 

We concluded that two main differences affect the improvement of the parallax error, which are (1) choice of maser feature for the parallax fit and (2) assignment of systematic error. Regarding the first difference, we choose two maser features (see Table \ref{table2} in Appendix \ref{appendix:1}), while \citet{2019arXiv190909930K} adopted three maser features. This is because the eliminated maser feature showed a signature of blending  (i.e. several peaks in spectra as shown in Fig. 5 of \citealp{2019arXiv190909930K}). About the second difference, we estimated individual systematic errors for the two maser features (see Table \ref{table2}), while \citet{2019arXiv190909930K} assumed the constant systematic error for the three maser features. Table \ref{table2} shows that the systematic error varies for the two maser features. It indicates a common systematic error (e.g., tropospheric zenith delay residual) and an individual systematic error, for example due to the maser structure, dominate the parallax error as previously discussed by \citet{2015PASJ...67...65N}.

We confirmed that our parallax error is degraded from 23 to 33 $\mu$as if we apply the two differences used in \citet{2019arXiv190909930K} for our data. Finally, we assign our result summarized in Table \ref{table:1} for the best parallax and proper motion of G125.52+02.03.

\begin{table*}[tbp]
\caption{Vertical height and velocity for star-forming regions, traced by maser sources with Galactocentric distances of $>$ 7 kpc.}
\label{table:5}
\begin{center}
\small
\begin{tabular}{rrrrrrrrrrr}
\hline
\hline
Source		&$R$\footnotemark[$*$]	&$z$\footnotemark[$*$] \ \ \  &$W$\footnotemark[$*$] \ \ 	&Refs.& 	 &Source		&$R$\footnotemark[$*$]	&$z$\footnotemark[$*$] \ \ \  &$W$\footnotemark[$*$] \ \ 	&Refs.\\

(G)	&(kpc)&(pc)	\ \ \ &(km s$^{-1}$)	&	&	&(G) &	(kpc)&(pc)	\ \ \ &(km s$^{-1}$)			\\
  \hline
007.47+00.05	&	12.37$^{+3.39 }_{-2.54 }$	&	17$\pm$2	&	0$\pm$11	&	1	&	&	107.29+05.63		&	8.41$^{+0.03 }_{-0.02 }$	&	76$\pm$6 	&	11$\pm$5	&	1	\\
014.33-00.64	&	7.07$^{+0.11 }_{-0.14 }$	&	-13$\pm$1	&	-3$\pm$8	&	1	&	&	108.18+05.51		&	8.48$^{+0.01 }_{-0.01 }$	&	87$\pm$1 	&	-4$\pm$5	&	1	\\
031.24-00.11	&	7.50$^{+2.59 }_{-1.58 }$	&	-26$\pm$4	&	5$\pm$12	&	1	&	&	108.20+00.58		&	10.40$^{+0.59 }_{-0.40 }$	&	44$\pm$7 	&	10$\pm$11	&	1	\\
040.42+00.70	&	8.47$^{+2.12 }_{-1.34 }$	&	156$\pm$26	&	14$\pm$6	&	1	&	&	108.42+00.89		&	9.25$^{+0.20 }_{-0.15 }$	&	38$\pm$4 	&	0$\pm$6	&	1	\\
040.62-00.13	&	8.25$^{+3.75 }_{-1.77 }$	&	-29$\pm$7	&	-3$\pm$9	&	1	&	&	108.47-02.81		&	9.67$^{+0.07 }_{-0.06 }$	&	-159$\pm$5 	&	-9$\pm$7	&	1	\\
042.03+00.19	&	9.71$^{+2.48 }_{-1.60 }$	&	46$\pm$7	&	-24$\pm$8	&	1	&	&	108.59+00.49		&	9.24$^{+0.12 }_{-0.10 }$	&	21$\pm$1 	&	0$\pm$5	&	1	\\
043.16+00.01	&	7.60$^{+0.67 }_{-0.52 }$	&	1$\pm$0	&	11$\pm$11	&	1	&	&	109.87+02.11		&	8.46$^{+0.01 }_{-0.01 }$	&	29	&	-1$\pm$5	&	1	\\
048.60+00.02	&	8.13$^{+0.42 }_{-0.35 }$	&	3$\pm$0	&	6$\pm$7	&	1	&	&	110.19+02.47		&	9.72$^{+0.59 }_{-0.35 }$	&	137$\pm$30 	&	14$\pm$12	&	1	\\
049.26+00.31	&	7.11$^{+0.83 }_{-0.48 }$	&	47$\pm$6	&	-7$\pm$7	&	1	&	&	111.23-01.23		&	9.86$^{+0.83 }_{-0.44 }$	&	-72$\pm$19 	&	-2$\pm$10	&	1	\\
058.77+00.64	&	7.02$^{+0.06 }_{-0.05 }$	&	37$\pm$4	&	-2$\pm$6	&	1	&	&	111.25-00.76		&	10.01$^{+0.13 }_{-0.12 }$	&	-48$\pm$2 	&	-3$\pm$6	&	1	\\
059.47-00.18	&	7.38$^{+0.03 }_{-0.03 }$	&	-6$\pm$0	&	-8$\pm$11	&	1	&	&	111.54+00.77		&	9.45$^{+0.07 }_{-0.07 }$	&	35$\pm$1 	&	-11$\pm$5	&	1	\\
059.78+00.06	&	7.31$^{+0.02 }_{-0.03 }$	&	2$\pm$0	&	-5$\pm$5	&	1	&	&	115.05-00.04		&	9.68$^{+0.23 }_{-0.18 }$	&	-2	&	-5$\pm$11	&	1	\\
059.83+00.67	&	7.05 	&	48$\pm$2	&	-2$\pm$5	&	1	&	&	121.29+00.65		&	8.67$^{+0.02 }_{-0.02 }$	&	10	&	-3$\pm$5	&	1	\\
060.57-00.18	&	8.28$^{+0.66 }_{-0.43 }$	&	-26$\pm$3	&	7$\pm$6	&	1	&	&	122.01-07.08		&	9.47$^{+0.07 }_{-0.06 }$	&	-268$\pm$11 	&	3$\pm$5	&	1	\\
069.54-00.97	&	7.65 	&	-42$\pm$1	&	6$\pm$5	&	1	&	&	123.06-06.30		&	9.96$^{+0.19 }_{-0.16 }$	&	-310$\pm$26 	&	-17$\pm$10	&	1	\\
070.18+01.74	&	8.93$^{+0.46 }_{-0.33 }$	&	223$\pm$22	&	-4$\pm$6	&	1, 2	&	&	123.06-06.30		&	9.64$^{+0.09 }_{-0.08 }$	&	-261$\pm$13 	&	-8$\pm$3	&	1	\\
071.31+00.82	&	8.00$^{+0.21 }_{-0.12 }$	&	67$\pm$8	&	-11$\pm$15	&	1	&	&	125.52+02.03		&	13.39$^{+1.14 }_{-0.81 }$	&	244$\pm$38 	&	-9$\pm$9	&	4, 5, d	\\
071.52-00.38	&	7.80$^{+0.06 }_{-0.03 }$	&	-24$\pm$2	&	-4$\pm$5	&	1	&	&	133.94+01.06		&	9.61$^{+0.03 }_{-0.03 }$	&	36	&	1$\pm$3	&	1	\\
073.65+00.19	&	13.53$^{+4.18 }_{-2.17 }$	&	44$\pm$11	&	-20$\pm$14	&	1	&	&	134.62-02.19		&	10.00$^{+0.08 }_{-0.08 }$	&	-93$\pm$3 	&	-6$\pm$6	&	1	\\
074.03-01.71	&	7.86 	&	-48$\pm$1	&	10$\pm$10	&	1	&	&	135.27+02.79		&	13.09$^{+0.38 }_{-0.33 }$	&	291$\pm$19 	&	1$\pm$10	&	1	\\
074.56+00.84	&	7.88$^{+0.10 }_{-0.02 }$	&	39$\pm$9	&	8$\pm$11	&	1	&	&	136.84+01.16		&	9.92$^{+0.73 }_{-0.40 }$	&	45$\pm$12 	&	4$\pm$6	&	1, 3	\\
075.29+01.32	&	10.67$^{+0.66 }_{-0.51 }$	&	213$\pm$19	&	-18$\pm$6	&	1	&	&	160.14+03.15		&	12.08$^{+0.10 }_{-0.10 }$	&	225$\pm$5 	&	5$\pm$6	&	1	\\
075.76+00.33	&	8.04$^{+0.06 }_{-0.04 }$	&	20$\pm$1	&	6$\pm$10	&	1	&	&	168.06+00.82		&	13.43$^{+0.71 }_{-0.56 }$	&	76$\pm$9 	&	5$\pm$8	&	1	\\
075.78+00.34	&	8.11$^{+0.13 }_{-0.08 }$	&	22$\pm$2	&	1$\pm$10	&	1	&	&	170.65-00.24		&	10.01$^{+0.31 }_{-0.23 }$	&	-8$\pm$1 	&	-7$\pm$10	&	1	\\
076.38-00.61	&	7.95$^{+0.01 }_{-0.01 }$	&	-14$\pm$0	&	13$\pm$19	&	1	&	&	173.48+02.44		&	9.82$^{+0.04 }_{-0.04 }$	&	71$\pm$1 	&	2$\pm$5	&	1	\\
078.12+03.63	&	7.98 	&	98$\pm$4	&	23$\pm$6	&	1	&	&				174.20-00.07		&	10.28$^{+0.10 }_{-0.09 }$	&	-3	&	9$\pm$6	&	1	\\
078.88+00.70	&	8.19$^{+0.07 }_{-0.05 }$	&	40$\pm$3	&	-22$\pm$11	&	1	&	&	176.51+00.20		&	9.11$^{+0.02 }_{-0.02 }$	&	3	&	0$\pm$5	&	1	\\
079.73+00.99	&	8.02 	&	23$\pm$1	&	6$\pm$6	&	1	&	&					182.67-03.26		&	14.51$^{+2.32 }_{-1.34 }$	&	-363$\pm$96 	&	11$\pm$11	&	1	\\
079.87+01.17	&	8.02 	&	32$\pm$1	&	4$\pm$10	&	1	&	&				183.72-03.66		&	9.73$^{+0.03 }_{-0.03 }$	&	-102$\pm$1 	&	4$\pm$10	&	1	\\
080.79-01.92	&	8.05$^{+0.01 }_{-0.00 }$	&	-55$\pm$4	&	-4$\pm$5	&	1	&	&	188.79+01.03		&	10.15$^{+0.52 }_{-0.34 }$	&	36$\pm$7 	&	-11$\pm$6	&	1	\\
080.86+00.38	&	8.05 	&	9$\pm$0	&	5$\pm$5	&	1	&	&						188.94+00.88		&	10.23$^{+0.03 }_{-0.03 }$	&	32	&	-4$\pm$6	&	1	\\
081.75+00.59	&	8.07 	&	15$\pm$0	&	7$\pm$3	&	1	&	&					192.16-03.81		&	9.63$^{+0.10 }_{-0.08 }$	&	-101$\pm$6 	&	4$\pm$6	&	1	\\
081.87+00.78	&	8.07 	&	17$\pm$0	&	2$\pm$3	&	1	&	&					192.60-00.04		&	9.78$^{+0.11 }_{-0.10 }$	&	-2	&	5$\pm$5	&	1	\\
090.21+02.32	&	8.18 	&	27$\pm$0	&	7$\pm$5	&	1	&	&					196.45-01.67		&	12.17$^{+0.19 }_{-0.18 }$	&	-121$\pm$5 	&	3$\pm$5	&	1	\\
090.92+01.48	&	10.11$^{+0.82 }_{-0.50 }$	&	151$\pm$27	&	17$\pm$7	&	1	&	&	209.00-19.38		&	8.49 	&	-138$\pm$1 	&	5$\pm$6	&	1	\\
092.67+03.07	&	8.39$^{+0.01 }_{-0.01 }$	&	87$\pm$2	&	-1$\pm$11	&	1	&	&	211.59+01.05		&	11.92$^{+0.17 }_{-0.16 }$	&	76$\pm$3 	&	-5$\pm$13	&	1	\\
094.60-01.79	&	9.36$^{+0.18 }_{-0.14 }$	&	-125$\pm$9	&	15$\pm$4	&	1, 3	&	&	213.70-12.59		&	8.87$^{+0.02 }_{-0.02 }$	&	-190$\pm$4 	&	1$\pm$5	&	1	\\
095.05+03.97	&	12.85$^{+2.02 }_{-1.21 }$	&	641$\pm$136	&	-2$\pm$8	&	4, a	&	&	217.79+01.05		&	13.53$^{+0.62 }_{-0.51 }$	&	112$\pm$11 	&	-4$\pm$7	&	1	\\	
095.29-00.93	&	9.86$^{+0.22 }_{-0.18 }$	&	-79$\pm$5	&	1$\pm$8	&	1	&	&		229.57+00.15		&	11.66$^{+0.23 }_{-0.20 }$	&	12	&	-10$\pm$15	&	1	\\
097.53+03.18	&	11.34$^{+0.57 }_{-0.44 }$	&	382$\pm$39	&	6$\pm$5	&	4, b	&	&		232.62+00.99		&	9.26$^{+0.08 }_{-0.07 }$	&	28$\pm$1 	&	1$\pm$4	&	1	\\
098.03+01.44	&	8.95$^{+1.12 }_{-0.33 }$	&	68$\pm$30	&	27$\pm$16	&	1	&	&	236.81+01.98		&	10.16$^{+0.20 }_{-0.17 }$	&	105$\pm$8 	&	-2$\pm$6	&	1	\\
100.37-03.57	&	9.41$^{+0.11 }_{-0.09 }$	&	-216$\pm$11	&	3$\pm$10	&	1	&	&	239.35-05.06		&	8.80$^{+0.05 }_{-0.04 }$	&	-104$\pm$6 	&	-3$\pm$3	&	1	\\
102.35+03.64	&	11.45$^{+0.76 }_{-0.55 }$	&	412$\pm$56	&	-3$\pm$10	&	4, c	&	&	240.31+00.07		&	11.76$^{+0.35 }_{-0.30 }$	&	6	&	6$\pm$12	&	1	\\
105.41+09.87	&	8.42$^{+0.02 }_{-0.02 }$	&	151$\pm$8	&	-13$\pm$5	&	1	&	&	\\


\hline

\multicolumn{10}{@{}l@{}}{\hbox to 0pt{\parbox{170mm}{\footnotesize
\par\noindent \\
Columns 1-2 show source name and Galactocentric distance, respectively. Columns 3 and 4 represent height from the Galactic mid-plane and velocity perpendicular to the disk in the direction of the north Galactic pole, respectively. Numbers and alphabets in column 5 show references for VLBI astrometric results and $V_{{\rm LSR}}$, respectively (see below). Note that the local standard of rest (LSR) velocity of the central star that excites the maser was estimated from observations of thermal line emission (e.g., CO) from the parent cloud. Columns 6-10 are the same as columns 5-10.
  \\
\footnotemark[$*$] For the calculations, A5 model of \citet{2019arXiv191003357R} was adopted: the Galactic constant  $R_{{\rm 0}}$ of 8.15 kpc; solar motion $W_{\odot}$ of 7.6 km s$^{-1}$. Note that a model of Galactic rotation curve, the Galactic constant $\Theta_{0}$, and the solar motion components ($U_{\odot}$, $V_{\odot}$) are not used when calculating the $W$ value. Error of $W$ is calculated, based on Appendix 1 of \citet{2015PASJ...67...69S} (see also Appendix \ref{appendix:2} of this paper). \\
$\bf References$: (1) \citet{2019arXiv191003357R}; (2) \citet{2019AJ....157..200Z}; (3) \citet{2019ApJ...876...30S}; (4) This paper;  (5) \citet{2019arXiv190909930K}; (a) \citet{2007PASJ...59.1185S}; (b) \citet{2010ApJS..188..313W}; (c) \citet{1993A&AS...98..589W}; (d) \citet{2003A&A...399.1083K}.
 \\
}\hss}}

\end{tabular}
\end{center}
\end{table*}

\section{Discussion}
Using our new results along with previous VLBI astrometric results in the literature (Table \ref{table:5}), we discuss vertical structure and velocities perpendicular to the disk (i.e., $W$) for the Galactic warp. For that we restrict sources with Galactocentric distances of $>$ 7 kpc. For the transformation from heliocentric to Galactic flames we refer to \citet{2009ApJ...700..137R}. The error of $W$ is estimated by considering errors of parallax, proper-motion components and $V_{{\rm LSR}}$ as summarized in \citet{2015PASJ...67...69S}.\footnote{See Appendix \ref{appendix:2} of this paper.} The Galactic constant $R_{0}$ of 8.15 kpc and the solar motion $W_{\odot}$ of 7.6 km s$^{-1}$ (\citealp{2019arXiv191003357R}) are assumed in the following discussion. Note that assumptions of the other Galactic constant $\Theta_{0}$, solar motion components ($U_{\odot}$, $V_{\odot}$) and Galactic rotation curve are not necessary for the following discussion.

\subsection{Vertical structure of the Galactic outer disk}

Figure \ref{fig:2} (top) shows source's height perpendicular to the disk (i.e. $z$ in pc) in the Cartesian coordinates. Our new results are associated with the Outer arm at the Galactic longitude range of 95$^{\circ}$ $< l <$ 126$^{\circ}$. The new results clearly trace the Galactic warp reaching to 200$\sim$600 pc toward the north Galactic pole. We simply compare the maximum hight of the current VLBI results with previous results in Table \ref{table:6}. The Galactic warp traced by the VLBI result (i.e. maser) is consistent with those traced by H {\scriptsize I} and Red clump within error, although we should carefully estimate possible uncertainties of the previous results for more accurate comparison (e.g., uncertainty of the kinematic distance).  

Figure \ref{fig:3} (top left) represents the height as a function of Galactocentric distance. It follows well known fact that the maximum height of the Galactic warp is increased with Galactocentric distance. Locally we can see the NGC281 superbubble as an anomaly at ($X$, $Y$) $\sim$ (2, 10) kpc (see also previous discussions in \citealp{2008PASJ...60..975S}, \citealp{2009ApJ...694..192M}, and \citealp{2014PASJ...66....3S}). Maser sources associated with the NGC 281 superbubble represent a Galactic latitude of $< -$6$^{\circ}$ and are located below the Galactic mid-plane with $z$ $\sim$ $-$300 pc (see Table \ref{table:5}). 

\begin{table}[htbp]
\caption{Galactic warp tracers.}
\label{table:6}
\begin{tabular}{lrrrc}
\hline
\hline

Tracer	&$ R$	&$\beta$&$z$ &Refs.	 \\

	&(kpc)&(deg)&(pc)&				\\
  \hline
Maser				&12.85$^{+2.02 }_{-1.21}$	&45.9$^{+6.1 }_{-5.1}$	&641$\pm$136	&1\\
H {\scriptsize I}		&$\sim$13			&50			&$>$500		&2\\
Cepheid				&13					&50			&$\sim$300	&3	\\
Dust					&13					&50			&$\sim$2,000&4	\\
Red clump			&13					&50			&$\sim$680&5	\\

\hline

\multicolumn{5}{@{}l@{}}{\hbox to 0pt{\parbox{80mm}{\footnotesize
\par\noindent 
\\
Column 1 shows different tracers for the Galactic warp. Columns 2 and 3 represent Galactocentric distance and Galactocentric azimuth ($\beta$), respectively. The Sun is located toward $\beta$ = 0 (deg) and the direction of the Galactic rotation is positive in $\beta$. Column 4 displays the height of the Galactic warp. Column 5 shows references (see below).\\
$\bf References$: (1) This paper; (2) \citet{2003PASJ...55..191N}; (3) \citet{2019NatAs...3..320C}; (4) \citet{2001ApJ...556..181D}; (5) \citet{2002A&A...394..883L}.
 \\
}\hss}}

\end{tabular}
\end{table}

\subsection{Vertical velocities of the Galactic outer disk}
\subsubsection{Vertical bulk motion}

\begin{figure*}[tbhp] 
 \begin{center} 
     \includegraphics[scale=1.0]{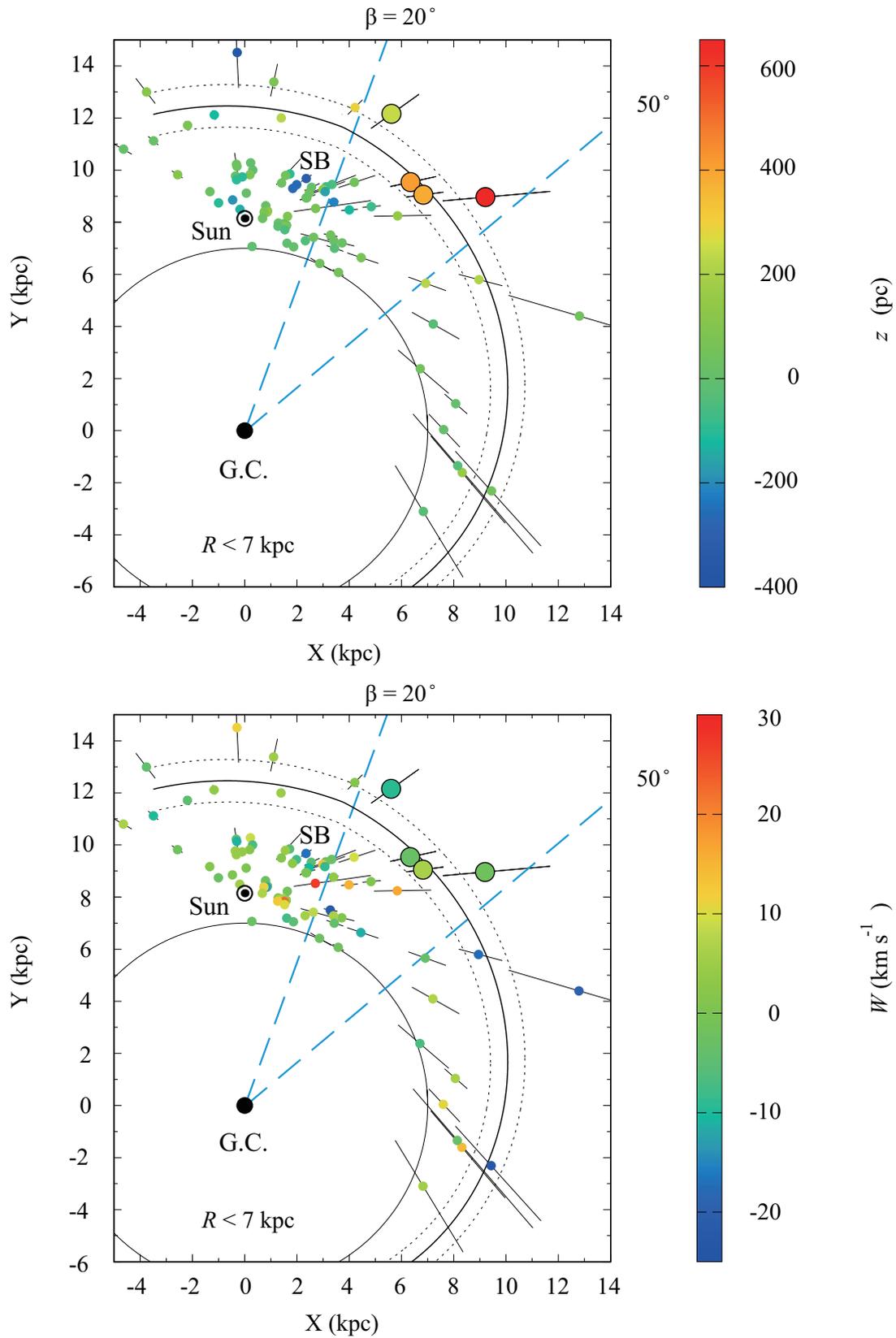} 
\end{center} 
\caption{ $(Top)$ Maser sources with Galactocentric distances of $>$ 7 kpc (i.e. Table \ref{table:5}) are projected on the Galactic plane as viewed from the north Galactic pole. The color indicates height perpendicular to the disk in the direction of the north Galactic pole (i.e. $z$ in pc). The Galactic center is at (0, 0) kpc and the solar position is assumed to be at (0, 8.15) kpc. Solid and dashed curves represent a logarithmic spiral-arm model and arm's width for the Outer arm, respectively (referring to \citealp{2019arXiv191003357R}). Large filled circles indicate results of this paper. The large circular arc indicates a distance of 7 kpc from the center. The label of ``$SB$'' means the location of the NGC281 superbubble. Cyan dashed lines represent Galactocentric azimuth $\beta$ (deg). The Sun is located toward $\beta$ = 0 (deg) and the direction of the Galactic rotation is positive in $\beta$.  $(Bottom)$ Same as (top), but for $W$ (km s$^{-1}$). $W$ is velocity perpendicular to the disk in the direction of the north Galactic pole. The solar motion $W_{\odot}$ of 7.6 km s$^{-1}$ (\citealp{2019arXiv191003357R}) is assumed for the calculation of $W$.}
\label{fig:2} 
\end{figure*}

Figure \ref{fig:2} (bottom) shows vertical velocity distribution for the VLBI astrometric results in the Cartesian coordinates. Positive $W$ at the Galactic anticenter of ($X$, $Y$) $\sim$ (0, 14.5) kpc is consistent with $Gaia$ DR2 results for young stars and giants (\citealp{2018MNRAS.481L..21P}). However, there is only one VLBI astrometric result with $R >$ 14 kpc, we will have to increase VLBI astrometric results around the region for confirming the validness of the positive $W$. If the VLBI astrometric result of the positive $W$ is confirmed, it can be regarded as the bending mode, where vertical bulk motion (i.e. positive $W$) is expected. The signature of the bending mode can be clear with increasing $R$ and toward the Galactic anticenter (e.g., see Fig. 2 of \citealp{2019MNRAS.490..797C}; Fig. 10 of \citealp{2019arXiv190511944W}). As previously discussed by \citet{2019MNRAS.490..797C}, the existence of the bending mode implies that the Milky Way has experienced external perturbations from e.g., a satellite galaxy or dark matter subhalos.

\begin{figure*}[tbhp] 
 \begin{center} 
     \includegraphics[scale=1.0]{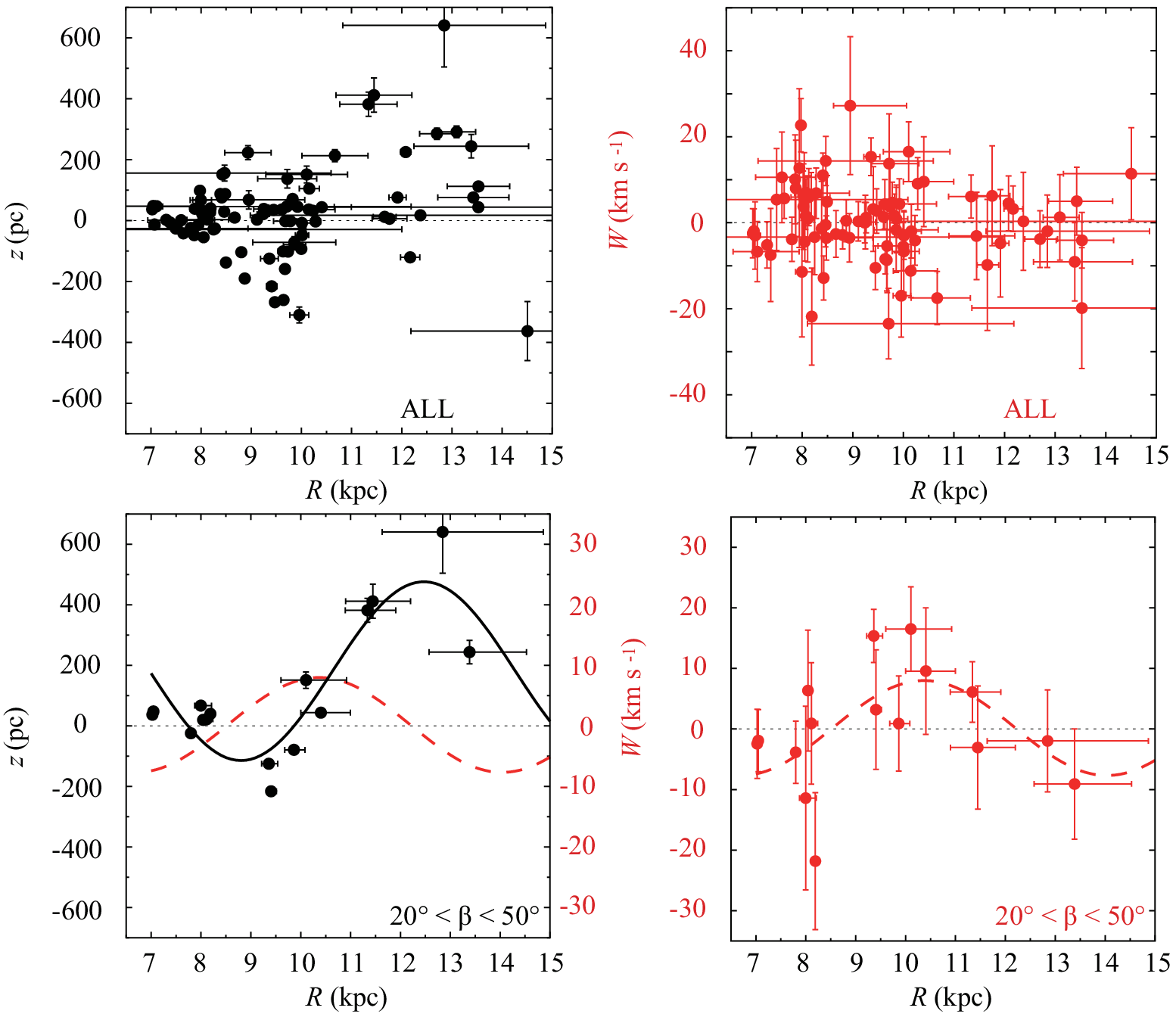} 
\end{center} 
\caption{ \textit{(Top left)} Galactic height $z$ (pc) is expressed as a function of Galactocentric distance ($R$ in kpc) for VLBI astrometric results (Table \ref{table:5}). \textit{(Top right)} Vertical velocity ($W$ in km s$^{-1}$) is expressed as a function of $R$ (kpc) for the results.
\textit{(Bottom left)} Same as (Top left), but for sources in the range of 20$^{\circ}$ $<$ Galactocentric azimuth ($\beta$) $<$ 50$^{\circ}$. Solid black curve shows a result of the unweighted least squares with $z$ = $a$ + b$\times$sin($R/c \times$2$\pi$) + d$\times$cos($R/c \times$2$\pi$), where $a$ = 181$\pm$35 (pc), $b$ = 281$\pm$94 (pc), $c$ = $-$7.3$\pm$0.7 (kpc) and $d$ = $-$91$\pm$223 (pc). Dashed red curve is copied from Bottom right figure for $W$ vs. $R$. Y2-axis represents $W$. Note that the both curves cannot be applied for sources with $R <$ 7 kpc (see the text). \textit{(Bottom right)} Same as (Bottom left), but for $W$. Dashed red curve shows a result of the unweighted least squares with $W$ = $e$ + f$\times$sin($R/(-7.3) \times$2$\pi$) + g$\times$cos($R/(-7.3) \times$2$\pi$), where $e$ = 0$\pm$2 (km s$^{-1}$), $f$ = $-$4$\pm$3 (km s$^{-1}$), and $g$ = $-$7$\pm$3 (km s$^{-1}$). The period in the sinusoids was fixed to be $-$7.3 kpc by referring to the result of $z$ vs. $R$.}
\label{fig:3} 
\end{figure*}  

\subsubsection{Oscillatory behavior}

To validate the oscillatory behavior between the vertical velocities (i.e. $W$ in km s$^{-1}$) and Galactic heights (i.e. $z$ in pc) as mentioned in the Section \ref{Intro}, we divided the VLBI astrometric results into arbitrary ranges of Galactocentric azimuth ($\beta$ in degrees), and found the clear signature of the oscillatory behavior in the range of 20$^{\circ}$ $<$ $\beta$ $<$ 50$^{\circ}$ (Figure \ref{fig:3} (bottom row)). In the $\beta$ range, $z$ value is maximized around $R \sim$ 12.5 kpc, at which $W$ value is around 0 km s$^{-1}$ as shown by black and red curves in Fig. \ref{fig:3} (bottom row). The black and red curves were obtained by the unweighted least squares , rather than the weighted least squares. The choice was motivated by two reasons that (1) distant source is down weighted in the weighted least squares and (2) the structure and kinematics of the Galactic warp is clearly traced by the distant source. 

Note that the both curves in Fig. \ref{fig:3} (bottom row) cannot be applied for VLBI astrometric results with $R <$ 7 kpc. This is because \citet{2019arXiv191003357R} revealed that the VLBI astrometric results within 7 kpc of the Galactic center have a scale height of only 19 $\pm$ 2 pc and these results are very tightly distributed in the Galactic plane.

The wave like behavior as a function of Galactocentric distance and the oscillatory behavior, as shown in Fig. \ref{fig:3} (bottom row), are consistent with the numerical simulation result of \citet{2013MNRAS.429..159G}, where the effect of the passage of the Sagittarius dwarf galaxy into the Milky Way was examined (see Fig. 6E in \citealp{2013MNRAS.429..159G}). Importantly, their simulation can also explain the inner flat disk, as mentioned above. Thus, the VLBI astrometric results suggests that the Galactic disk has experienced external perturbations from e.g., the Sagittarius dwarf galaxy.

\section{Summary}

We reported parallax and proper motion results for four 22 GHz water maser sources as part of VERA ORC project (see Table \ref{table:1} and Fig. \ref{fig:1}). These show Galactic latitudes of $>$ 2$^{\circ}$ and trace the Galactic warp reaching to 200$\sim$600 pc above the Galactic mid-plane at the Galactic longitude range of 95$^{\circ}$ $< l <$ 126$^\circ$ (see the large filled circles in Fig. \ref{fig:2}). 

New results along with previous results in the literature showed that the maximum height of the Galactic warp is increased with Galactocentric distance (Figures \ref{fig:2} (top) and \ref{fig:3} (top left)). Examining velocities perpendicular to the disk hinted at the existence of the bending mode as is the case for $Gaia$ DR2 results in \citet{2018MNRAS.481L..21P}.  

Also, we found wave like behaviors of the vertical velocity $W$ (km s$^{-1}$) and Galactic height $z$ (pc) in radial direction. There is the signature of oscillatory behavior between $W$ and $z$ (Fig. \ref{fig:3} (bottom row)), however, which must be confirmed by further observations. The both results suggest that the Milky Way has experienced external perturbation(s) from a perturbing satellite (e.g. the Sagittarius dwarf galaxy) as previously predicted by the numerical simulation of \citet{2013MNRAS.429..159G}.

The observational results summarized in this paper favor a scenario that the Milky Way has experienced external perturbations from satellite galaxies and/or dark matter subhalos.

\begin{ack}
We acknowledge the anonymous referee for valuable comments, which improved the manuscript.
We would like to thank VERA project members for supports of observations, correlation processing and data reductions. TH is financially supported by the MEXT/JSPS KAKENHI Grant Number 17K05398.
\end{ack}




\bibliographystyle{apj}
\bibliography{reference}


\appendix
\onecolumn
\section{Supplemental materials}\label{appendix:1}


\begin{table*}[htbp]
\caption{Trigonometric parallaxes for individual sources\footnotemark[$*$].}
\label{table2}
\begin{center}
\begin{tabular}{cccccccc}
\hline
\hline
Source&Feature	&$V_{\rm{LSR}}$	&Detection	&Parallax	&	&\multicolumn{2}{c}{Errors	}\\
 \cline{7-8} 
&		&km s$^{-1}$	&in Epoch\footnotemark[$\dag$]							&(mas)	&	&R.A.	&Decl.	\\
&		&	&				&				&	&\multicolumn{2}{c}{(mas)}	\\
\hline
G095.05+3.97 &1a	&-78.7$\sim$-77.0		&$\times$$\times$$\times$$\times$GHIJKLM		&0.102$\pm$0.022		&	&0.052		&0.043		\\
			&1b	&-82.1$\sim$-78.3		&$\times$$\times$$\times$$\times$GHIJKLM		&0.117$\pm$0.026		&	&0.046		&0.132		\\
\multicolumn{3}{l}{Combined fit}							&&0.108$\pm$0.016	&		&	&		\\
\multicolumn{3}{l}{Final result}							&&0.108$\pm$0.023	&		&	&		\\
\hline
\\
G097.53+3.18&1	&-71.3$\sim$-70.4		&ABCDEFGHIJK		&0.177$\pm$0.028		&	&0.092		&0.070		\\
\hline
\\

G102.35+3.64&1	&-86.7$\sim$-85.0		&NOPQRSTUVWXY		&0.154$\pm$0.021		&	&0.047		&0.134		\\
\hline
\\

G125.51+2.03&1	&-50.8$\sim$-48.3		&abcdefghijkl		&0.147$\pm$0.030		&	&0.083		&0.091		\\
			&2	&-61.8$\sim$-59.2		&$\times$$\times$$\times$$\times$efghijkl		&0.144$\pm$0.020		&	&0.032		&0.073		\\
\multicolumn{3}{l}{Combined fit}							&&0.145$\pm$0.016	&		&	&		\\
\multicolumn{3}{l}{Final result}							&&0.145$\pm$0.023	&		&	&		\\
\hline

\hline

\multicolumn{4}{@{}l@{}}{\hbox to 0pt{\parbox{140mm}{\footnotesize
\par\noindent
\\
\footnotemark[$*$] Positional errors were adjusted so that the reduced chi-square becomes unity. The brightest maser spot was used for the parallax fit for each feature. Since maser feature's data could be correlated due to similar differential atmospheric delay differences between maser features and a background QSO, we inflated parallax error of combined fit by $\sqrt[]{N_{\rm{feat}}}$, where $\sqrt[]{N_{\rm{feat}}}$ is the number of maser features used for the parallax fit.\\
\footnotemark[$\dag$] Please refer to Table \ref{table:4} for the corresponding epoch date. \\
}\hss}}
\end{tabular}
\end{center}
\end{table*}

\begin{table*}[htbp]
\caption{Systematic proper motions for individual sources.}
\label{table3}
\begin{center}
\begin{tabular}{ccccc}
\hline
\hline
Source&Feature	&	&\multicolumn{2}{c}{Proper Motion\footnotemark[$\ast$]}	\\
\cline{4-5}
&	&$V_{\rm{LSR}}$	&$\mu_{\alpha} \rm{cos}\delta$	&$\mu_{\delta}$	\\
&	&km s$^{-1}$	&(mas yr$^{-1}$)&(mas yr$^{-1}$)	\\
\hline
G095.05+3.97&1a$-$1c			&-85.9 $\sim$ -77.0 	&-1.93$\pm$0.05	&-2.73$\pm$0.04		\\
&2			&-81.5 $\sim$ -80.6	&-2.71$\pm$0.19	&-2.91$\pm$0.17		\\
&3			&-85.5 $\sim$ -85.1 	&-2.53$\pm$0.08	&-2.52$\pm$0.07		\\
&4			&-87.4 $\sim$ -86.7 	&-2.61$\pm$0.03	&-2.35$\pm$0.03		\\
\hline
&Unweighted mean		&-83.3				 	&-2.44$\pm$0.18\footnotemark[$\dag$]	&-2.63$\pm$0.12\footnotemark[$\dag$]		\\ \\

G097.53+3.18&1			&-71.3 $\sim$ -70.4 	&-2.37$\pm$0.03	&-2.56$\pm$0.05		\\
&2a$-$2c			&-78.9 $\sim$ -76.1	&-2.96$\pm$0.04	&-2.56$\pm$0.07		\\
&3			&-76.5 $\sim$ -75.5 	&-2.71$\pm$0.03	&-2.96$\pm$0.05		\\
&4a$-$4d			&-76.7 $\sim$ -74.2 	&-2.74$\pm$0.07	&-1.88$\pm$0.12		\\
&5			&-85.0 $\sim$ -82.2	&-2.96$\pm$0.21	&-2.21$\pm$0.36		\\
&6			&-79.1 	&-2.66$\pm$0.17	&-2.09$\pm$0.28		\\
&7a$-$7b	&-80.1 $\sim$ -78.2 	&-2.48$\pm$0.03	&-2.67$\pm$0.06		\\
&8			&-78.2 $\sim$ -77.6 	&-2.29$\pm$0.14	&-2.45$\pm$0.24		\\
&9			&-79.4 $\sim$ -74.0 	&-2.57$\pm$0.06	&-2.02$\pm$0.10		\\

\hline
&Unweighted mean		&-77.0				 	&-2.64$\pm$0.08\footnotemark[$\dag$]	&-2.38$\pm$0.12\footnotemark[$\dag$]		\\ \\

G102.35+3.64&1			&-86.7 $\sim$ -85.0	&-2.53$\pm$0.02	&-2.14$\pm$0.04		\\
\hline \\

G125.51+2.03&1			&-50.8 $\sim$ -48.3  	&-1.51$\pm$0.04	&-0.06$\pm$0.02		\\
&2			&-61.8 $\sim$ -59.2	&-0.87$\pm$0.07	&0.39$\pm$0.04		\\
&3			&-52.9 $\sim$ -51.2 	&-0.85$\pm$0.12	&-0.89$\pm$0.07		\\
&4a$-$4b	&-59.2 $\sim$ -55.9 	&-1.49$\pm$0.08	&-0.47$\pm$0.05		\\
&5a$-$5b	&-63.4 $\sim$ -59.2	&-1.29$\pm$0.05	&-0.61$\pm$0.03		\\ 
\hline
&Unweighted mean		&-56.3				 	&-1.20$\pm$0.14\footnotemark[$\dag$]	&-0.33$\pm$0.22\footnotemark[$\dag$]		\\ 

\hline

\multicolumn{4}{@{}l@{}}{\hbox to 0pt{\parbox{120mm}{\footnotesize
\par\noindent \\
\footnotemark[$\ast$]
Proper motion components in east ($\alpha$cos$\sigma$) and north ($\sigma$) directions were determined by adapting a final parallax (see Table \ref{table2}).  \\
\footnotemark[$\dag$]
Systematic proper motion components, regarded as motion components of the central star that excites the masers, were determined by unweighted average of the proper motion components. Errors in the systematic proper motion components mean the standard error. \\

}\hss}}

\end{tabular}
\end{center}
\end{table*}

\begin{figure*}[htbp] 
 \begin{center} 
     \includegraphics[scale=1.15]{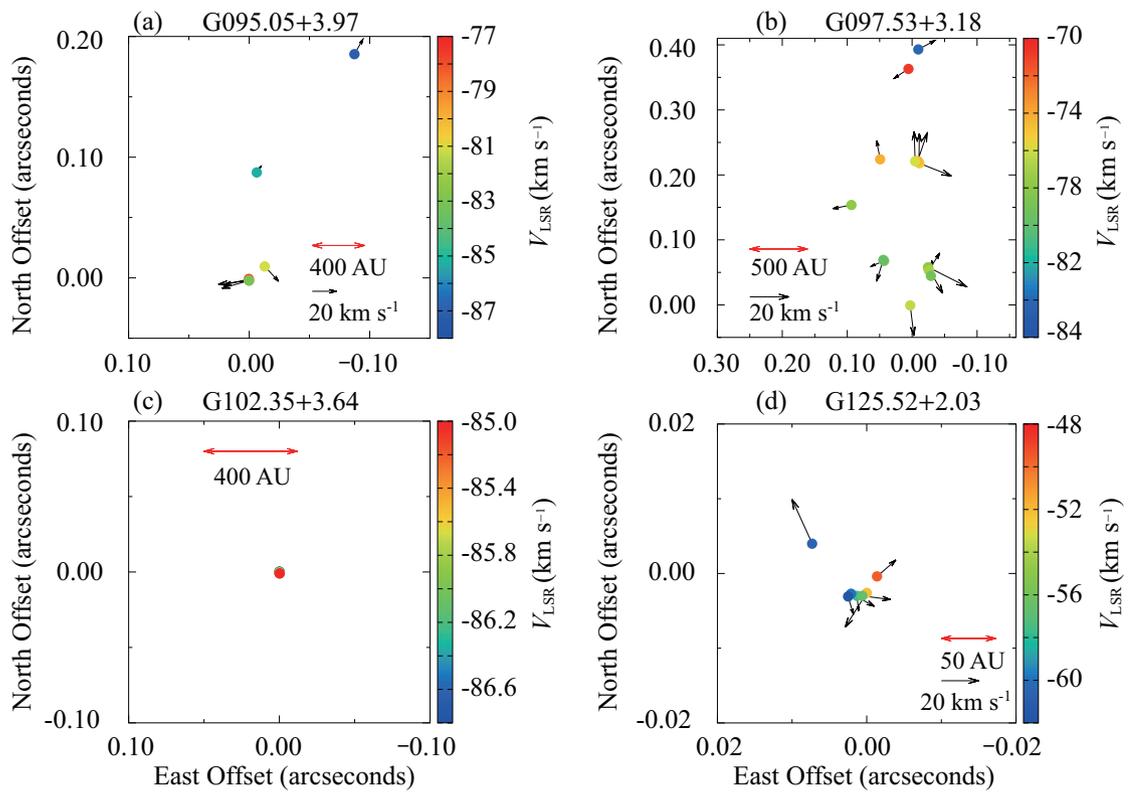} 
\end{center} 
    \caption{Maser-spot and internal-motion distributions for {\bf (a)} G095.05+3.97, {\bf(b)} G097.53+3.18,
{\bf(c)} G102.35+3.64, and {\bf(d)} G125.52+2.03. 
  The origin of coordinates for each map is described in Table \ref{table:4}. The internal motion vectors were drawn by subtracting systematic proper motion components from each maser spot motion (see Table \ref{table3}).
The horizontal red and black arrows in each map show
an absolute spatial and velocity scales, converted at a source distance (see Table \ref{table:1}), respectively. 
Color bar indicates the local standard of
rest (LSR) velocity.}    
     \label{fig:5} 
\end{figure*}


\onecolumn
\section{Erratum for Appendix 1 of \citet{2015PASJ...67...69S}}
\label{appendix:2}

We have found two mistakes in Appendix 1 of  \citet{2015PASJ...67...69S}, and thus we show the corrections here. First, the sign of the cross product in the transformation matrix $T$ of \citet{2015PASJ...67...69S} should be dropped as 

\[
      \bf{T} = 
\left[
    \begin{array}{lll}
      +\rm{cos}\theta_{0}	&+\rm{sin}\theta_{0}&0\\
      +\rm{sin}\theta_{0}	&-\rm{cos}\theta_{0}&0\\
      0						&0					&+1
    \end{array}
    \right]
\left[
    \begin{array}{lll}
          -\rm{sin}\delta_{\rm{NGP}}	&0	&+\rm{cos}\delta_{\rm{NGP}}\\
          0								&-1	&0\\
          +\rm{cos}\delta_{\rm{NGP}}	&0	&+\rm{sin}\delta_{\rm{NGP}}\\
    \end{array}
  \right]
\left[
    \begin{array}{lll}
          +\rm{cos}\alpha_{\rm{NGP}}	&+\rm{sin}\alpha_{\rm{NGP}}	&0\\
          +\rm{sin}\alpha_{\rm{NGP}}	&-\rm{cos}\alpha_{\rm{NGP}}	&0\\
          0								&0							&+1\\
    \end{array}
  \right]
\]

where $\alpha_{\rm{NGP}}$ and $\delta_{\rm{NGP}}$ are right ascension and declination of the north Galactic pole in J2000 coordinates (see \citealp{2009ApJ...700..137R}). Zero of Galactic longitude is at position angle $\theta_{0}$ = 122\fdg932 which is angle of separation between the equatorial plane and the Galactic plane.

Second, non-circular motion components of ($U$, $V$, $W$) at source's position should be described as

\[
\left[
    \begin{array}{c}
      U \\
      V \\
      W 
    \end{array}  
\right]
=
\bf{C} \cdot 
\left[
    \begin{array}{c}
      v_{\rm{helio}}\\
      k\mu_{\alpha}\rm{cos}\delta/\pi \\
      k\mu_{\delta}/\pi
    \end{array}
  \right]
  + \bf{D} \cdot
  \left[
    \begin{array}{c}
      U_{\odot}\\
      \bf{\Theta_{0}}{\rm +}{\it V}_{\odot}\\
      W_{\odot}
    \end{array}
  \right]
  -
  \left[
    \begin{array}{c}
      0\\
      \Theta(R)\\
      0
    \end{array}
  \right]
  \]
where the Galactic constant ($\Theta_{0}$), as represented by the bold font, is inserted into the original formula of \citet{2015PASJ...67...69S} to calibrate the relative observables from the Sun. The observables are heliocentric velocity ($v_{{\rm helio}}$), trigonometric parallax ($\pi$), and proper motion components of ($\mu_{\alpha}$cos$\delta$, $\mu_{\delta}$) in the directions of right ascension and declination. The conversion factor $k$ is 4.74057. Conversion matrices $\bf{C}$ and $\bf{D}$ are defined in \citet{2015PASJ...67...69S}. Solar motion components of $U_{\odot}$, $V_{\odot}$ and $W_{\odot}$ are directed toward the Galactic center, Galactic rotation and the north Galactic pole, respectively. $\Theta(R)$ shows a rotation curve. 



\end{document}